\documentclass{vgtc}                          % final (conference style)
\graphicspath{{figures/}{pictures/}{images/}{./}} % where to search for the images

\usepackage{times}                     % we use Times as the main font
\usepackage{tabu}                      % only used for the table example
\usepackage{booktabs}                  % only used for the table example
\usepackage{lipsum}                    % used to generate placeholder text
\usepackage{mwe}                       % used to generate placeholder figures

\usepackage{mathptmx}                  % use matching math font
\usepackage{comment}
\usepackage{placeins}

\usepackage{tikz}

\onlineid{0}

\vgtccategory{Research}

\vgtcinsertpkg

\title{Influence of Interactivity in Shaping User Experience and Social Acceptability of Mobile Augmented Reality}

\author{
 Tanja Koji\'c$^1$, Maurizio Vergari$^1$, Maximilian Warsinke$^1$, Sebastian M\"oller$^{1,2}$,  Jan-Niklas Voigt-Antons$^3$ \\
 $^1$Quality and Usability Lab, Technical University of Berlin, Germany\\
 $^2$German Research Center for Artificial Intelligence (DFKI), Berlin, Germany\\
 $^3$Immersive Reality Lab, Hamm-Lippstadt University of Applied Sciences, Germany
}

\abstract{
    This study investigates the impact of the Degree of Interactivity on User Experience (UX) and social acceptability (SA) in Mobile Augmented Reality (MAR) applications. As AR technologies become more prevalent, understanding how varying levels of interactivity influence both user perception and social dynamics is crucial for their design and adoption. Two commercially available MAR applications, IKEA and Virtlo, which differ significantly in their interactivity levels, were used to conduct a user study. The study examines how body movements required for interaction with AR content affect both UX and SA, shedding light on users' comfort levels and potential social barriers in public settings. The findings suggest a complex relationship between interactivity, perceived usability, and social considerations, emphasizing the need for a balanced design approach. This research provides valuable insights into the development of future AR applications by addressing not only usability but also the broader social implications of AR interactions. By integrating social acceptability into traditional UX evaluations, this study highlights its significance in ensuring the seamless integration of AR technologies into everyday environments.
} % end of abstract


\begin{document}



\maketitle

\newcommand\copyrighttext{%
    \footnotesize \textcopyright 2026 IEEE. Personal use
    of this material is permitted. Permission from IEEE
    must be obtained for all other uses, in any current or
    future media, including reprinting/republishing this
    material for advertising or promotional purposes,
    creating new collective works, for resale or
    redistribution to servers or lists, or reuse of any
    copyrighted component of this work in other works.
    https://doi.org/10.1109/VRW66409.2025.00114}

\newcommand\copyrightnotice{%
\begin{tikzpicture}[remember picture,overlay,shift=
    {(current page.south)}]
    \node[anchor=south,yshift=10pt] at (0,0)
    {\fbox{\parbox{\dimexpr\textwidth-\fboxsep-
    \fboxrule\relax}{\copyrighttext}}};
\end{tikzpicture}%
}
\copyrightnotice

\section{Introduction \& Related Work}
The last decade has witnessed a significant shift in the technological landscape, marked by rapid advancements that have not only introduced new products and services but also redefined the way users interact with digital content. Augmented Reality (AR), a technology that overlays computer-generated enhancements onto real-world objects, has emerged as a key innovation, creating hybrid experiences that seamlessly blend virtual and physical environments \cite{azuma1997survey}. This transformative technology has demonstrated its versatility across a wide range of industries, including marketing, healthcare, manufacturing, and entertainment, where it enhances engagement and interaction in various domains.

One of AR’s most impactful applications is in educational settings, where its ability to create interactive and immersive experiences has been shown to significantly enhance learning and training environments \cite{wu2013current}. By providing real-time visualizations, simulations, and hands-on engagement, AR bridges the gap between theoretical knowledge and practical application, making complex concepts more accessible and fostering deeper understanding. In addition to structured learning environments, AR has also gained traction in professional training scenarios, such as medical simulations, engineering prototyping, and skill-based vocational training, where precision and contextual interaction are crucial.

In parallel with AR's growing prominence, Mobile Augmented Reality (MAR) applications have experienced exponential growth, driven by the widespread adoption of smartphones and tablets equipped with increasingly powerful sensors, processors, and cameras. These portable devices enable users to access AR experiences on the go, making the technology more accessible, adaptable, and integrated into everyday life. From navigating urban spaces with augmented directions to virtually trying on clothes or placing digital furniture, MAR applications offer a diverse range of use cases that make them highly relevant in both private and public settings. This increased accessibility has also led innovation in areas such as tourism, retail, and social media, where MAR applications contribute to enriched experiences of user engagement.

However, while the growing presence of MAR applications offers numerous opportunities, it also brings forth a set of challenges that affect both User Experience (UX) and social acceptability (SA). The interactive nature of AR often requires users to perform distinct gestures, movements, or device manipulations, which may not always align with conventional social norms in public spaces. As AR applications become more integrated into everyday interactions, it is essential to examine how different levels of interactivity influence user perception, usability, and the extent to which these interactions are socially acceptable in various contexts. Addressing these challenges is crucial to ensuring the seamless adoption of AR technologies and optimizing their design for both usability and social comfort.

Despite the fact that MAR has a wide range of applications and is increasingly being incorporated into everyday situations, the interaction dynamics that occur in public or shared spaces present a number of difficult challenges. As an example, the presence of bystanders has the potential to significantly impact both UX and SA of Augmented Reality technologies. In this context, User Experience is defined as the perceptions and responses of the user that are the result of the use of a technology or the anticipated use of a technology \cite{ketola2008exploring}. Meanwhile, social acceptability refers to how the technology is perceived from the standpoint of comfort or discomfort by both users and observers, reflecting a critical aspect of technology adoption and integration in social settings \cite{alallah2018performer}\cite{distler2018acceptability}.

While previous research has focused extensively on the usability and technical capabilities of MAR applications, this study uniquely examines the intersection of UX and SA. By integrating these dimensions, we provide a comprehensive understanding of how interactivity levels shape both the personal and social dynamics of MAR usage, addressing a critical gap in existing literature. Previous studies have explored usability aspects of MAR, but fewer have directly examined how interactivity levels influence both UX and SA in real-world environments. This study aims to fill this gap by investigating how user engagement and social comfort are affected by different interaction modalities in MAR applications.

Research has shown that one of the most critical factors influencing the success of MAR applications is the degree of interactivity they require from users. Interactivity in AR can range from minimal, such as simple touch gestures, to highly immersive, requiring full-body movement and constant user engagement. While higher interactivity may enhance the immersive qualities of an experience, it can also lead to greater social discomfort, particularly in public spaces where users may feel self-conscious about their actions. This balancing act between increasing user engagement through interactivity and maintaining a socially acceptable experience forms a core challenge in AR design. 

Moreover, the demographic characteristics of users, including age and gender, play a significant role in how MAR applications are perceived and adopted. Younger users, who are often more familiar with mobile technology, may approach AR with fewer barriers, while older users might find the interactive demands of these applications more challenging or intrusive. Similarly, gender differences have been observed in how users perceive social visibility when interacting with AR in public spaces. Women, for instance, may be more concerned with how their actions are perceived by others, affecting their comfort with using interactive AR technologies in certain environments.

MAR applications are characterised by the degree of physical engagement that is required from users. The purpose of this paper is to investigate the impact that the Degree of Interactivity in MAR applications has on both UX and SA. By examining how different levels of interactivity affect these aspects, this study aims to contribute insights into the possible design and implementation of AR applications across two environments.

\subsection{User Experience in MAR}

Research in UX of MAR applications is relatively new, with most scientific contributions emerging in the recent decade as MAR are becoming closer to mainstream technology. Dirin and Laine \cite{dirin2018user} argue that traditional mobile UX design principles may still apply to MAR, though the latter demands more from users in terms of mental and physical engagement. This includes a complete focus on the applications and often involves using both hands, engaging both short-term and long-term memory. Their studies highlight several challenges unique to MAR UX, such as battery drain, substantial processing requirements, screen size, full-body interaction, and the absence of adequate prototyping tools \cite{dirin2018user}. They emphasize the importance of designing MAR applications that are sensitive to the user's environmental awareness and promote emotional engagement through satisfying and user-friendly interactions.

Irshad and Rambli \cite{irshad2014user} further classify MAR UX into three domains: as a phenomenon, a field of study, and a practice, endorsing User-Centered Design (UCD) as a methodology for developing MAR applications. Building on this sentiment, Olsson et al. \cite{olsson2013expected} stress that understanding users' fundamental needs is crucial for the successful adoption of MAR applications.

\subsection{Presence, Immersion, and Emotions}

The constructs of Presence and Immersion in virtual environments vary across studies but generally refer to the psychological sense of "being there" and the technological capability to deliver a vivid virtual experience, respectively \cite{slater1997framework}. Although this study does not streach into these dimensions, this paper adopts Slater and Wilbur's definitions as a foundation \cite{slater1997framework}. The emotional responses elicited by MAR and other virtual technologies often utilize dimensional models to categorize emotions into valence, arousal, and dominance, as initially proposed by Mehrabian \cite{mehrabian1970semantic} and Russell \cite{russell1980circumplex}, providing a framework for assessing user emotions during interaction with virtual content.

\subsection{Social Acceptability}

The social acceptability of a technology, which is important for its success, reflects its perceived appropriateness by users and observers within a social context. Alallah et al. 
\cite{alallah2018performer} define it as the level of comfort or discomfort experienced by both performers and spectators when interacting with or observing the technology. Despite growing interest, significant research opportunities remain in understanding the social dynamics of immersive technologies \cite{alallah2018performer}.

From the performer’s perspective, studies by Rico and Brewster \cite{rico2010gesture}\cite{rico2010usable} indicate that device-based gestures are often seen as more socially acceptable than body-based gestures, as devices provide a context that justifies the user's actions in public spaces. Similarly, Tung et al. \cite{tung2015user} found a preference for less conspicuous interaction modalities, a sentiment echoed by Ahlström et al. \cite{ahlstrom2014you}, who noted that smaller, quicker gestures are favored for their lower social obtrusiveness.

From the observer’s perspective, the social acceptability of AR is less studied but significant. Profita et al. \cite{profita2013don} observed that cultural and gender differences impact how technology usage is perceived by others. Denning et al. \cite{denning2014situ} highlighted that observers often have neutral or negative reactions to public use of AR, a point of consideration for designers aiming to integrate these technologies into socially sensitive environments. Schwind et al. \cite{schwind2018virtual} and Hsieh et al. \cite{hsieh2016designing} further argued that the social setting significantly influences the acceptability of AR, noting that usage in more interactive social environments like cafes may face greater scrutiny and resistance.

\subsection{Objectives} 

The objectives of this research include determining which characteristics and factors lead to increased MAR adoption in social settings and contributing to the improvement of design guidelines and principles to improve the UX of MAR applications. The general idea of existing scientific work is that less visible input modalities and gestures have a higher likelihood of social acceptability than noticeable movements for any technology. However, due to a lack of research on the relationship between the level of interactivity with UX and the social acceptability of AR and MAR in particular, this paper intends to answer the following research questions:

\begin{itemize}
\item How does the level of interactivity affect UX and social acceptability of MAR?
\item Do demographic characteristics such as age, and gender affect UX and social acceptability of MAR?
\end{itemize}

The degree of interactivity in mobile AR applications varies widely, ranging from simple, passive interactions (such as swiping and tapping) to more dynamic, immersive experiences that require physical movement (such as rotating the device or interacting with 360-degree views). Previous research suggests that higher levels of interactivity, while potentially more engaging, can also increase user discomfort in public settings where social visibility is a factor. In particular, the physical nature of these interactions can make users more self-conscious, affecting their overall comfort and social acceptability. This research seeks to clarify whether higher interactivity leads to improved UX due to enhanced engagement or reduced UX and SA due to increased physical demands and social scrutiny in public spaces.

User demographics, particularly age and gender, have been shown to influence how technologies like AR are adopted and experienced. Younger users, often more familiar with mobile technology and its interfaces, may find AR interactions more intuitive and socially acceptable than older users, who may perceive them as intrusive or physically demanding. Gender differences have also been documented in technology use, with women potentially being more conscious of social visibility in public settings, affecting their comfort levels with interactive AR apps. Men, on the other hand, may prioritize the immersive qualities of AR experiences over concerns about social acceptability. By examining these demographic factors, this research aims to provide insights into how age and gender influence the UX and SA of MAR applications, offering valuable guidance for the design of future AR experiences.

Existing research highlights that demographic characteristics such as age and gender significantly influence the adoption and perception of emerging technologies. Younger users, often more familiar with mobile technology interfaces, may find AR interactions more intuitive and less socially intrusive, whereas older users might perceive higher interactivity levels as physically or cognitively demanding. Additionally, gender differences have been observed, with women more frequently reporting concerns about social visibility and public scrutiny during technology use. These differences suggest that age and gender could play crucial roles in shaping both UX and SA for MAR applications, warranting their inclusion as focal points of this study.

%%%%%%%%%%%%%%%%%%%%%%%%%%%%%%%%%%%%%%%%%%%%%%%%%
\section{METHODS}
\subsection{Study Design and Setup}
The experiment was conducted on the [removed for double blind], using an iPhone 14 Pro Max to access and interact with two selected MAR applications: the IKEA app and Virtlo.
These apps were chosen due to their differing levels of interactivity. The IKEA app, which requires minimal user interaction—such as tapping, swiping, and basic pointing of the camera—is used to design a leisure or study space by placing 3D models in a real environment. Virtlo, on the other hand, demands more dynamic interaction by requiring the user to engage in a 360-degree panorama to locate places of interest in a city, thereby increasing the degree of interactivity involved. IKEA and Virtlo were selected due to their popularity, widespread adoption, and distinct interactivity levels, ensuring a meaningful comparison of user engagement and social acceptability. Other MAR apps, such as Amazon and Lego, were excluded due to differences in primary use cases or hardware constraints.
Also the two MAR applications, IKEA and Virtlo, were selected for this study due to their contrasting levels of interactivity, allowing us to investigate how different interaction demands influence UX and SA.

\begin{itemize}
    \item IKEA App: This application requires minimal interaction, primarily limited to tapping, swiping, and pointing the camera to place 3D furniture models in real environments. These simple interactions make the user’s experience relatively passive, as the app involves little physical engagement. Such low interactivity is ideal for public use, as it minimizes user movement and reduces social visibility.
    \item Virtlo App: In contrast, Virtlo demands a higher degree of interactivity. Users must engage with the environment by rotating their devices to explore a 360-degree panorama and locate points of interest in a cityscape. This higher interactivity level requires continuous user input and more noticeable physical actions. These interactions increase user engagement but may also heighten the user’s awareness of bystanders and the social context, potentially impacting social acceptability.
\end{itemize}

Justification for out selection of applications is that we selected IKEA and Virtlo to examine how varying levels of interactivity affect UX and SA in mobile AR. Although the applications serve different purposes—IKEA for interior design and Virtlo for outdoor POI navigation—both provide interactive AR experiences, making them suitable for comparison.
IKEA offers high interactivity through detailed manipulation of 3D objects in an indoor environment, while Virtlo focuses on navigation and location-based tasks in an outdoor context. These differences in task complexity and user control allow us to explore how interactivity influences UX and SA across distinct domains.

\begin{figure}[tb]
 \centering 
 \includegraphics[width=\columnwidth]{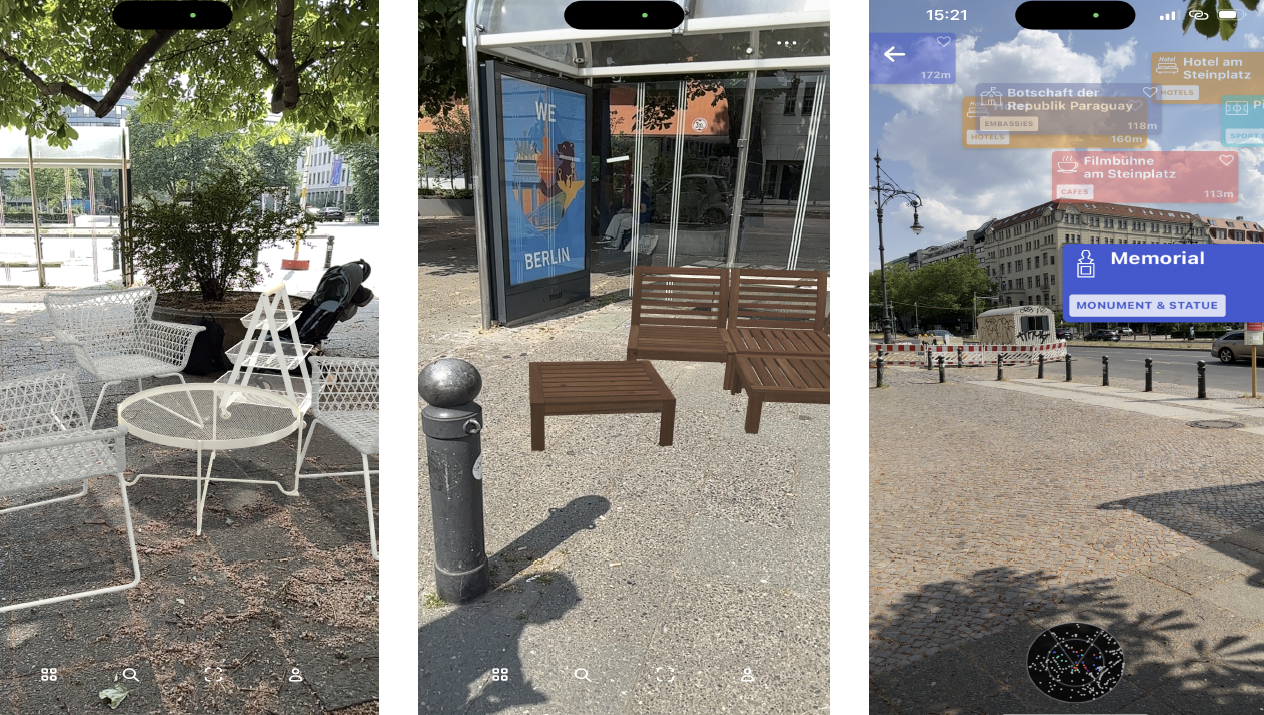}
 %\caption{Detailed Screenshots Illustrating the Range of Tasks Completed by Study Participants Using the IKEA and Virtlo Applications. }
 \caption{Comparison of MAR application interfaces. The left and middle images represent the IKEA app, showcasing its object-based interactivity for furniture placement. The right image represents the Virtlo app, illustrating its navigation-based Augmented Reality experience.}
 \label{fig:apps}
\end{figure}

By examining two applications as it is illustrated with Figure \ref{fig:apps} with such differing levels of interactivity, this study seeks to understand how the degree of physical engagement affects both the user’s comfort and social visibility, providing valuable insights into the design considerations for future AR applications.
Understanding these dynamics is crucial for ensuring that AR experiences are both engaging and seamlessly integrated into everyday interactions without causing discomfort or social friction.

% To ensure the table is positioned at the top of the next page across both columns
\FloatBarrier
\begin{table*}[tp]
\centering
\caption{Results of Wilcoxon Signed Rank Test Combined Table of All Significant Results for Social Accetabiliy and User Experience parameters}
\label{tab:combined_significant_results}
\begin{tabular}{lccc}
\toprule
\textbf{Description}                    & \textbf{Parameter}                 & \textbf{Z-score} & \textbf{Significance (p-value)} \\
\midrule
Interactivity - Social Acceptability    & Concerned of physical collision             & -2.220           & \textbf{0.026}                  \\
Interactivity - User Experience         &  Complicated – Easy                        & -2.126           & \textbf{0.033}                  \\
Gender Differences - Social Acceptability & Interesting if the others see what I am seeing in the app (Virtlo)     & -2.539           & \textbf{0.011}                  \\
Gender Differences - Social Acceptability & Concerned of physical collision (IKEA)     & -2.000           & \textbf{0.046}                  \\               
\bottomrule
\end{tabular}
\end{table*}

\subsection{Test Procedure}
Participants were individually scheduled for the study to avoid any interaction between them. Upon arrival, each participant was briefed by a moderator who explained the study's purpose and the functionality of the apps involved. After completing a pre-experiment questionnaire that gathered demographic data and previous AR experience, participants were introduced to the specific interactions required for each application. They engaged with the apps in a predetermined sequence to ensure consistency across the study, with each session randomly alternating the order of app usage to control for order effects.

Following the interaction with the MAR applications, participants completed several post-experiment questionnaires designed to assess various psychological and experiential aspects of their interaction. The igroup Presence Questionnaire (IPQ) was used to assess the sense of "being there," measuring the level of presence experienced in the virtual environment \cite{slater1997framework}. The Short User Experience Questionnaire (UEQ-S) measures the pragmatic and hedonic quality of the User Experience, providing insights into the practical and emotional aspects of using the MAR applications \cite{schrepp2017design}. The Self-Assessment Manikin (SAM) evaluates emotional responses across three dimensions—pleasure, arousal, and dominance \cite{bradley1994measuring}. The Social Acceptability Questionnaire (SAQ) was adapted from its original Virtual Reality (VR) context to evaluate the social acceptability of MAR applications from the user's perspective, focusing on their comfort and the perceived appropriateness of their actions \cite{eghbali2018social}.

All questionnaires used in this study are well-established, standardized instruments, ensuring the reliability and validity of the data collected.

\subsection{Participants}
A total of 20 participants (11 male and 9 female) were recruited, with ages ranging from 20 to 45 years old. The majority had some level of familiarity with AR technologies, although their experience varied from novice to adept. Most participants in the study were from the IT sector, but to ensure a broader perspective, individuals from diverse fields such as PR/Marketing, medicine, and business were also included. This varied sample was selected to provide a more comprehensive understanding of the usability and social acceptability of the MAR applications examined in the study, capturing insights from users with different professional backgrounds and experiences.

%%%%%%%%%%%%%%%%%%%%%%%%%%%%%%%%%%%%%%%%%%%%%%%%%%%%%%%%%%%%%%%%%%%%%%%%%%%%%%%%
\section{RESULTS}

To analyze the results, several statistical analyses were conducted to explore the relationships between demographic factors and UX and SA ratings. These analyses provided deeper insights into how interactivity, age, gender, and AR experience influence the perceived usability of the IKEA and Virtlo applications. Specifically, the Wilcoxon Signed Rank Test was used to detect statistically significant differences between conditions, comparing UX and SA ratings across different levels of interactivity. Additionally, descriptive statistics and visualizations were employed to illustrate trends in user responses. Table 1 provides an overview of the significant effects found, including metrics for social acceptability and User Experience.

\begin{figure}[h!]
 \centering 
 \includegraphics[width=\columnwidth]{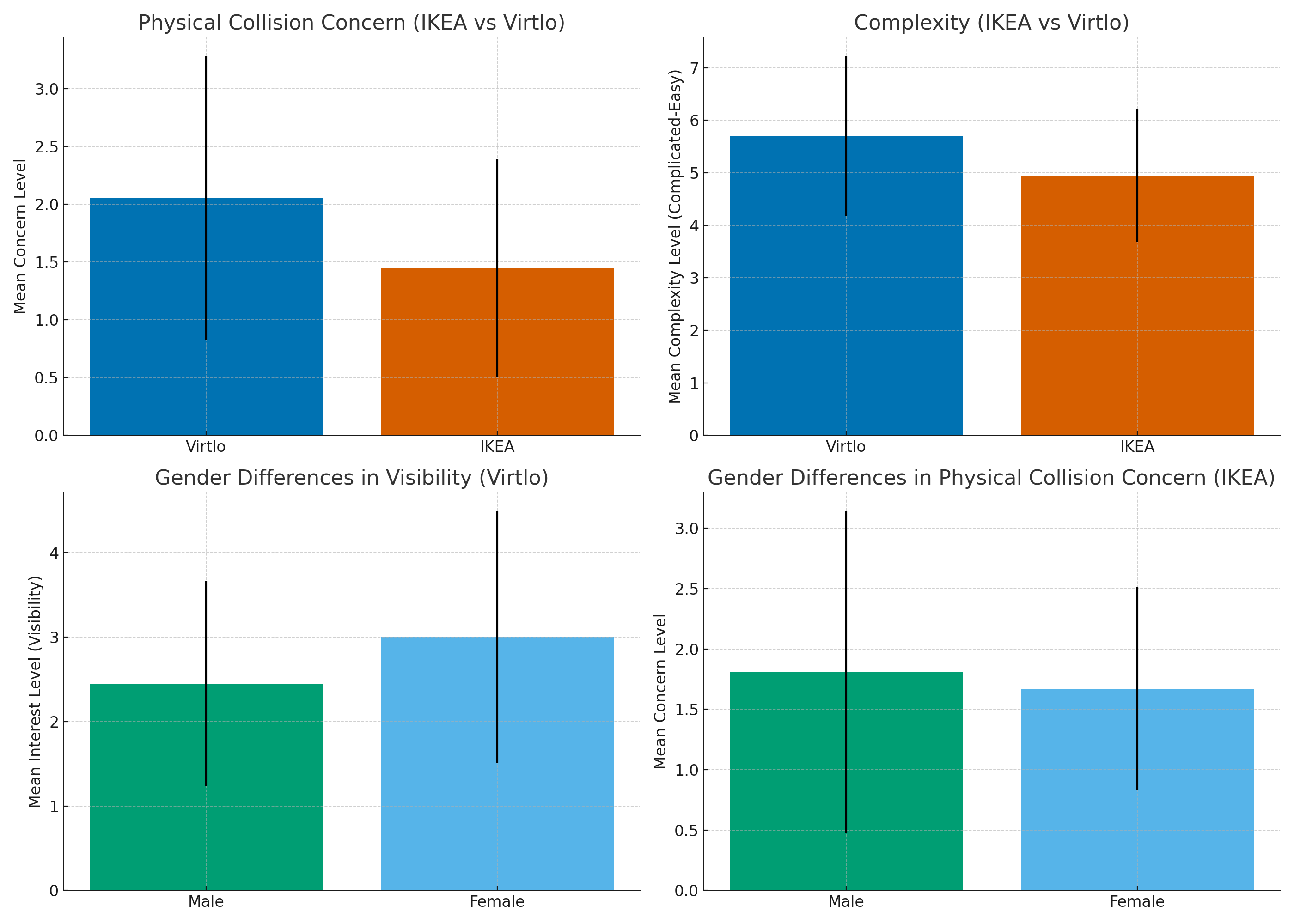}
 \caption{This figure presents four comparisons of UX and SA metrics based on significant results. The top-left subplot compares users' concerns about physical collisions between the IKEA and Virtlo apps. The top-right subplot illustrates the perceived complexity (Complicated-Easy) of both apps. The bottom-left subplot shows gender differences in interest regarding whether others can see what the user is seeing. The bottom-right subplot highlights gender differences in concerns about physical collisions.}
 \label{fig:graphs0}
 \vspace{-1em}
\end{figure}

\subsection{Interactivity and Demographic Impact Analysis}

Test indicated a clear concern among users about the potential of physical collision with objects when interacting with the Virtlo app (M = 2.05, SD = 1.23), which demands greater interactivity, in contrast to the IKEA app (M = 1.45, SD = 0.94). However, when assessing other social acceptability factors such as appropriateness, rudeness, comfort, visibility of actions to others, and privacy concerns related to recording, no other statistically significant differences were observed between the two apps. The only significant distinction in User Experience due to interactivity levels arose in the context of perceived complexity. Participants reported that Virtlo (M = 5.70, SD = 1.52), which entails more active engagement, was notably easier to use than IKEA (M = 4.95, SD = 1.27). Figure \ref{fig:graphs0} illustrates these significant differences, highlighting that users were more concerned about potential physical collisions with Virtlo, while also perceiving it as more engaging but more complex compared to IKEA.

Analysis revealed that female users were significantly more interested if the other people could see what I was doing and seeing in the app (M = 3.00, SD = 1.49) when using the Virtlo app compared to their male counterparts (M = 2.45, SD = 1.22). No other substantial gender-based differences were detected in social acceptability metrics. For the IKEA app, men showed a higher concern about accidental collisions (M = 1.81, SD = 1.33) than women (M = 1.67, SD = 0.84), which was statistically significant. A pronounced divergence was discovered in how men and women experienced the Virtlo app; women rated the interface more favorably overall. Men described Virtlo as more inefficient, dull, uninteresting, and mundane relative to women’s assessments.

Participants were categorized into two age groups—those in their twenties and those in their thirties—to determine the influence of age on social acceptability. This categorization did not result any significant differences in the social acceptability ratings.

\subsection{Comparison of UX Metrics Between IKEA and Virtlo}

\begin{figure}[h!]
 \centering 
 \includegraphics[width=\columnwidth]{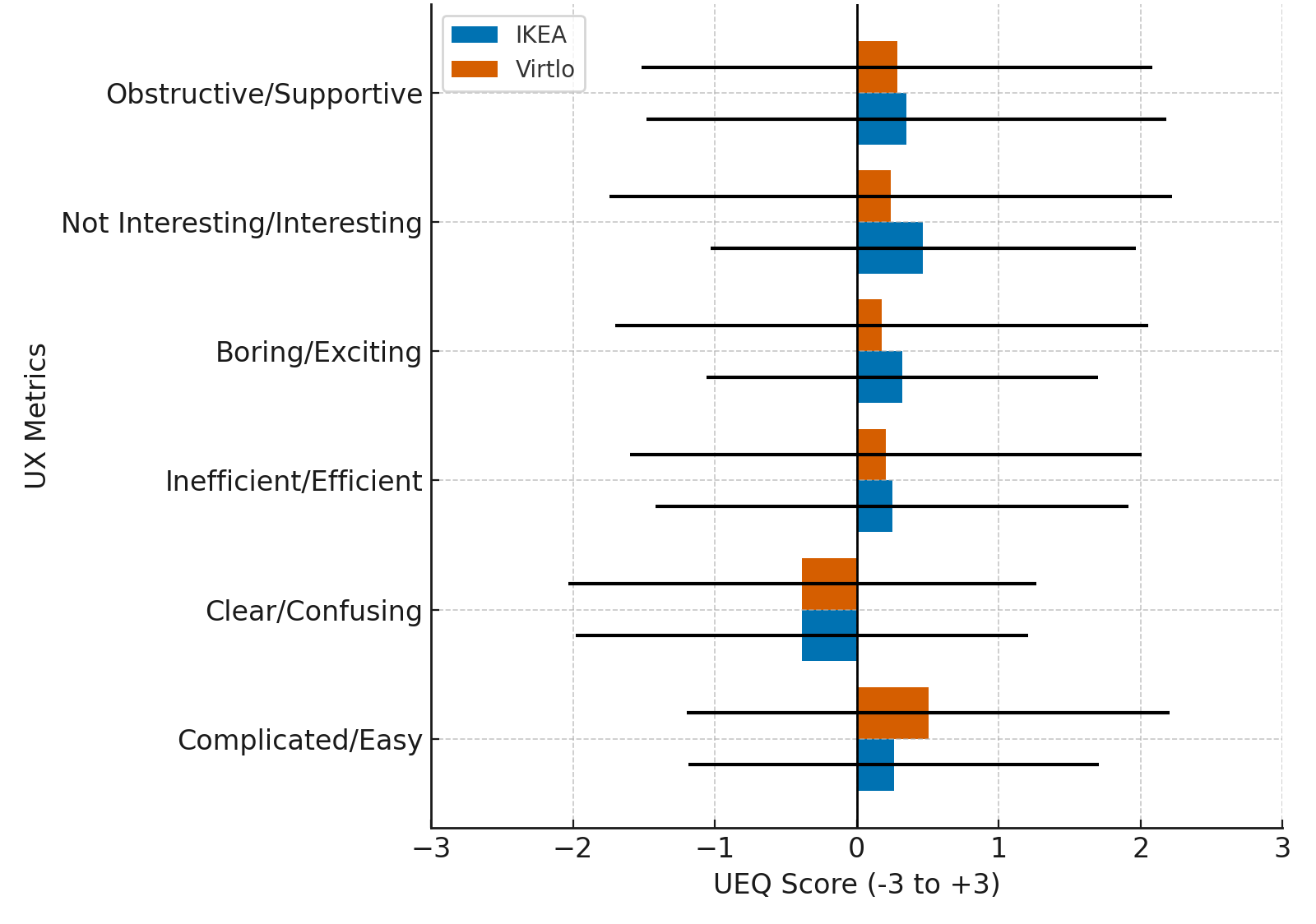}
 \caption{Comparison of UX metrics for the IKEA and Virtlo applications using UEQ scores on a scale from -3 to +3. Error bars show variability in the ratings.}
 \label{fig:graphs1}
 \vspace{1em}
\end{figure}

As illustrated in Figure \ref{fig:graphs1}, a comparison of the UX metrics between the IKEA and Virtlo applications revealed key differences in how participants perceived the two apps. Specifically, Virtlo, which requires higher interactivity, was rated as significantly more complicated (M = 5.70, SD = 1.52) and less clear (M = 4.95, SD = 1.27) compared to IKEA. Moreover, participants found Virtlo to be more engaging but also more obstructive in public settings.

\subsection{Gender Differences in UX and Social Acceptability}

\begin{figure}[h!]
 \centering 
 \includegraphics[width=\columnwidth]{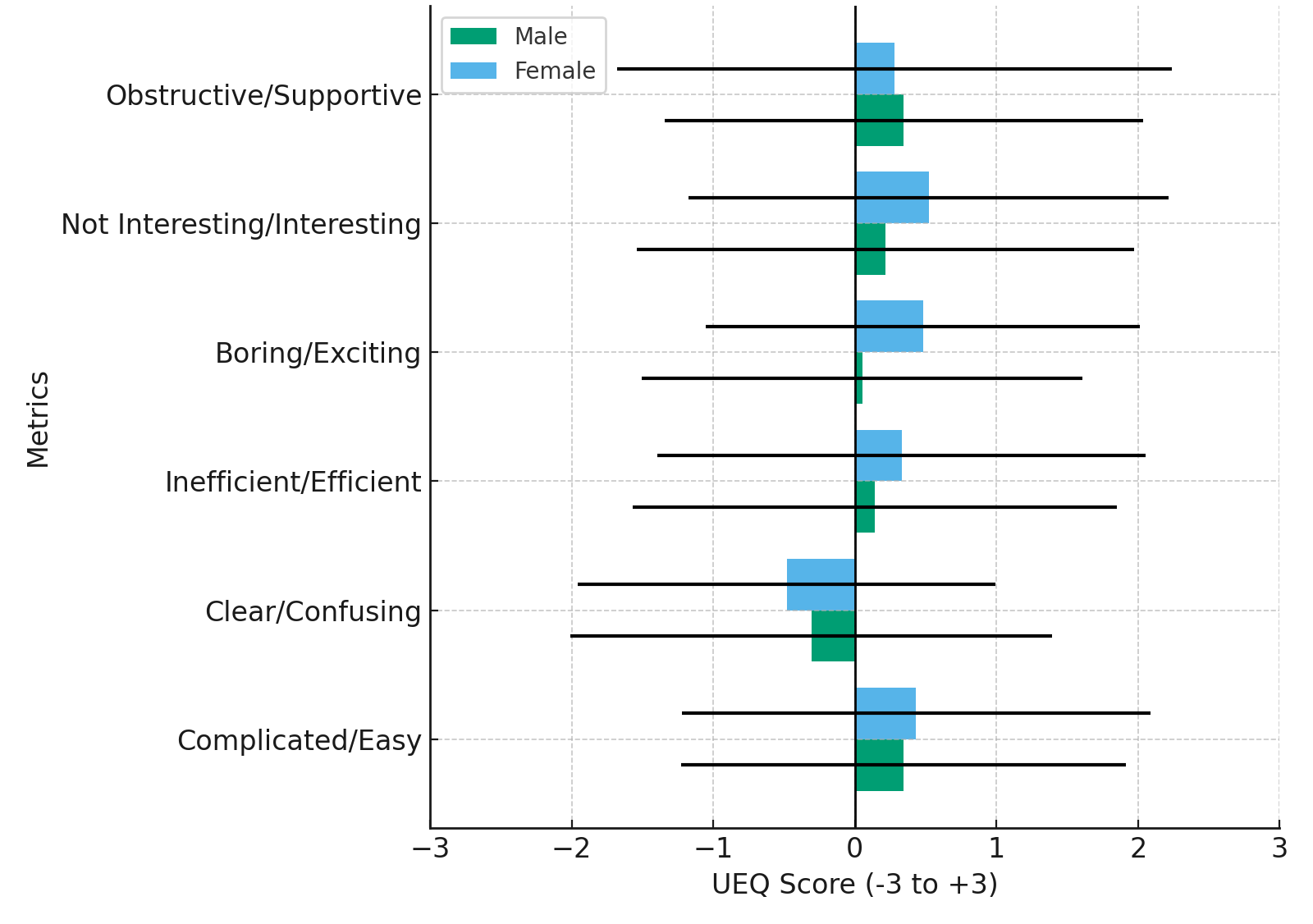}
 \caption{Comparison of UX metrics for gender differences using UEQ scores on a scale from -3 to +3. Error bars show variability in the ratings.}
 \label{fig:graphs2}
  \vspace{1em}
\end{figure}

Figure \ref{fig:graphs2} demonstrates that gender played a role in how participants rated UX and SA metrics across both applications. Female participants, on average, showed higher levels of concern regarding visibility of their actions (M = 3.00, SD = 1.49) and were more sensitive to social acceptability factors such as appropriateness. Male participants, on the other hand, tended to rate the apps as more inefficient (M = 6.00, SD = 1.00) and complicated (M = 7.00, SD = 1.50), particularly with the IKEA app.

\subsection{Correlation Between AR Experience and UX Metrics}

The correlation analysis between AR experience and UX metrics revealed weak but notable relationships. For instance, participants with more AR experience tended to feel more dominant (r = 0.25) and engaged (r = 0.26) when using the apps, suggesting that familiarity with AR technology increases comfort with its use. 

Again, there was little correlation between AR experience and perceptions of complexity, indicating that even experienced users can find certain aspects of AR interaction challenging.

\subsection{Impact of Age on UX and SA Metrics}

The analysis of UX and SA metrics across age groups highlighted some differences, particularly in the sense of presence ("Sense of Being There"). Older participants (30+) reported a higher sense of being immersed in the app environment (M = 6.00, SD = 1.00) compared to younger participants (M = 4.50, SD = 1.20). However, no significant differences were observed in other metrics such as efficiency or complexity.

\subsection{Comparison of Specific UX Factors Between IKEA and Virtlo by Gender}

Lastly, a gender-specific comparison of UX factors between the two applications. It reveals that female participants generally rated the Virtlo app as clearer (M = 5.70, SD = 1.20) and more engaging (M = 4.95, SD = 1.27), while male participants rated it as more inefficient (M = 6.00, SD = 1.00) and obstructive (M = 7.00, SD = 1.50). These findings suggest that gender plays a role not only in the social acceptability of AR apps but also in how users engage with interactive features.

\section{DISCUSSION}
This study aimed to investigate the impact of interactivity levels on UX and SA in Mobile Augmented Reality applications. The analyses provided insights into the role of demographic factors such as gender and age in shaping these perceptions, further contributing to our understanding of the complexities associated with MAR technology adoption.

%4.1 Interactivity Levels in IKEA and Virtlo Applications
The comparison between the IKEA and Virtlo applications revealed key differences in how users perceive interactivity. Virtlo, characterized by a higher level of physical interaction, was perceived as more complicated and inefficient than IKEA, which aligns with previous studies suggesting that higher interactivity can result in increased cognitive and physical load for users. However, Virtlo was also rated as more exciting and inventive, demonstrating that users value immersive, dynamic experiences despite the added complexity. These findings highlight the delicate balance between increasing interactivity to enhance user engagement and the need to ensure usability and efficiency. In the context of MAR, it is crucial for developers to consider how varying levels of interactivity can affect the overall experience and the user’s ability to comfortably engage with the technology, particularly in public spaces.

%4.2 Gender differences in SA and UX
This study shows rather interesting gender variations. Particularly when using Virtlo, the more interactive application, female participants voiced more worry about their actions being visible in public places. This is consistent with the body of current research implying women might be more sensitive to public perceptions of their activities. With IKEA scoring lower in these areas, male participants tended to concentrate more on the efficiency and usability of the applications. These results imply that gender influences not only users' assessment of the practicality of an application but also their social dynamic navigation of public interaction with technology. Ensuring larger adoption and satisfaction depends on designing AR applications that fit several points of view.

%4.3 Age and Immersion.
Examining age-related variations turned out some fascinating, if faint, patterns. Older participants especially with Virtlo reported a stronger sense of immersion. This would imply that the novelty and immersive nature of AR experiences have more influence on older users. Younger consumers, who probably know more about digital interactions, could find AR applications less original and, hence, less immersive. But age had no appreciable impact on views of complexity or efficiency, suggesting that interactivity levels—rather than age—have more bearing on determining these features of UX.

%4.4 AR Experience and UX
Unlike what would be expected, past experience with AR had no appreciable effect on opinions of complexity or inefficiency in either application. Even those who had used AR prior thought Virtlo to be difficult. This implies that even experienced users find difficulties when the complexity of interactions rises. Experience with AR technology does, however, help users feel more confident in navigating immersive environments; experienced AR users did report higher degrees of engagement and feelings of control. This result supports the need of streamlining user interactions to enable AR applications to be pleasant and accessible for any larger audience.

In alignment with previous studies on MAR usability, this research supports findings \cite{dirin2018user}, who highlighted the importance of designing user-friendly applications to minimize cognitive and physical load. However, unlike prior studies that primarily focused on functionality, this study explores the social acceptability of MAR applications, extending the discussion to public interaction dynamics. The observed gender differences in social visibility align with the previous related work \cite{rico2010usable}, yet our results highlight a need for further exploration of these differences in highly interactive environments. Moreover, while existing research has often emphasized younger users' familiarity with AR technology, this study adds a novel perspective by examining older users' immersion and engagement with interactive MAR applications. 

These findings align with prior research demonstrating that while higher interactivity can enhance engagement, it may also introduce cognitive load and social discomfort, particularly in public settings \cite{rico2010gesture, ahlstrom2014you}. Our results extend this discussion by showing that gender and age further modulate these effects, highlighting the need for adaptive MAR designs that account for different user preferences and sensitivities.

One limitation of this study is the relatively small participant pool, which may restrict the generalizability of the findings. A larger and more diverse sample would allow for a broader understanding of how demographic and contextual factors influence UX and SA in Mobile Augmented Reality applications. Future research should focus on expanding participant diversity to include a wider range of ages, cultural backgrounds, and familiarity with AR technologies, ensuring that the findings are applicable. % to more varied user populations.

\section{CONCLUSION}
This work offers a novel perspective of how age, gender, and interactivity affect the User Experience and social acceptability of Mobile Augmented Reality applications. Higher interactivity, can improve engagement and excitement but usually at the expense of higher complexity and inefficiency. Users seem willing to accept some degree of complexity in return for a more immersive experience, but it is important to make sure these interactions remain understandable and controllable, especially in public environments.

Gender variations were also noticeable; female participants were more worried about social visibility than male participants, who were instead concentrated more on the functional features of the applications. These results emphasise the need of gender-sensitive design strategies that take into account the several ways users prioritise social and functional components when interacting with AR technologies.
Though age did not show any clear influence on UX, older participants—especially in the more interactive application—reported higher feeling ofimmersion. This implies that AR apps may be new and more appealing to older users. Thus, developers have a chance to target this group with customised experiences.

Interestingly, past AR experience did not take enough into consideration the difficulties brought by more interactive applications. This underscores the need of creating straightforward, easily available AR experiences independent of user technological knowledge. Simplifying interactions without sacrificing engagement will be crucial to ensuring the broad usability and success of future AR applications.
While this study provides valuable insights, the relatively homogeneous participant pool may limit generalizability. Future research should include more diverse samples to explore broader user perspectives.

In conclusion, the results of this study highlight the need for careful consideration of both interactivity and demographic factors when designing MAR applications. Developers should strive to create experiences that are not only engaging but also usable and socially acceptable across a diverse range of users. Future research could further explore how specific design elements, such as reducing the complexity of interactions or enhancing social comfort, might improve the overall experience of using MAR applications in everyday settings.

\section{Acknowledgements}
We acknowledge the assistance from our colleagues and the project framework that enabled this research. This work would not have been possible also without the dedicated efforts of Anna Iliasova, whose contributions were involved in gathering the results presented in this paper.

\bibliographystyle{abbrv-doi}

\bibliography{main}
\end{document}